\documentclass[12pt]{article}
\usepackage{jheppub,epsfig,amssymb,amsmath,psfrag,subfigure}
\usepackage[utf8]{inputenc} 

\catcode`\@=11
\def \tr {\mathop{\rm tr}\nolimits}

\def \e  {\mathop{\rm e}\nolimits}
\newcommand\lr[1]{{\left({#1}\right)}}
\newcommand \widebar [1] {\overline{#1}}

\newcommand\re[1]{(\ref{#1})}

\def \qqqquad {\qquad\qquad}

\newcommand \vev [1] {\langle{#1}\rangle}
\newcommand{\ft}[2]{{\textstyle\frac{#1}{#2}}}

\def\numberbysection{\@addtoreset{equation}{section}
                     \def\theequation{\thesection.\arabic{equation}}}

\numberbysection

\title{Energy correlations in the end-point region}
 
\author {G.P.~Korchemsky}
 \affiliation {Institut de Physique Th\'eorique\footnote{Unit\'e Mixte de Recherche 3681 du CNRS}, Universit\'e Paris Saclay, CNRS, CEA, 91191 Gif-sur-Yvette}
\preprint{  \parbox[t]{28mm}{IPhT--T19/041}}

 \abstract
{
The energy-energy correlation (EEC) measures the angular distribution of the energy that flows  through two calorimeters separated by some relative angle  in the final state created by a source. We study this observable in the limit of small and large angles when it describes the correlation between particles belonging, respectively, to the same jet and to two almost back-to-back jets.  We present a new approach to resumming large logarithmically enhanced corrections in both limits that exploits the relation between the energy correlations and four-point correlation functions of conserved currents.  
At large angle, we derive the EEC from the behaviour of the correlation function in the limit when four operators are light-like separated in a sequential manner.  At small angle, in a conformal theory, we obtain 
the EEC from resummation of the conformal partial wave expansion of the correlation function at short-distance separation between the calorimeters. 
In both cases, we obtain a concise representation of the EEC in terms of the conformal data of twist-two operators and verify 
it by comparing with the results of explicit calculation at next-to-next-to-leading  order  in maximally supersymmetric Yang-Mills theory.
As a byproduct of our analysis, we predict the maximal weight part of the analogous QCD expression in the back-to-back limit.
}

\begin{document}
 
\maketitle

\flushbottom

\section{Introduction}\label{sect1}

Event shapes are important QCD observables describing the properties of the energy flow in hadronic final states of $e^+e^-$ annihilation. Their measurement at PEP, KEK, PETRA, SLD and LEP has improved our understanding of jets and allowed for a precise determination of the strong coupling constant.  

The energy-energy correlation (EEC) is one of the best studied event shapes. It measures a differential angular distribution of particles
 that flow through two calorimeters separated by the angle $0<\chi<\pi$ and is defined as an energy-weighted cross section 
corresponding to the process $e^+e ^-\to V \to a+ b+ \text{everything}$ (with $V$ being a virtual photon $\gamma^*$ or a $Z^0$ boson) \cite{Basham:1978bw,Basham:1978zq}
\begin{align}\label{EEC-def}
\text{EEC}(\chi) = \sum_{a,b} \int d\sigma_{V \to \, a+ b+ X} {E_a E_b\over Q^2}\delta(\cos\theta_{ab}-\cos\chi)\,,
\end{align}
where $E_a$ and $E_b$ are the energies of the detected particles,  $Q$ is the total centre-of-mass energy and the differential cross-section is normalized in such a way that $\sum_{a,b} \int d\sigma_{V \to \, a+ b+ X}=1$. The measurements of the EEC  yield  a smooth positive definite function of the angle  $0<\chi<\pi$  that is picked around the end points \cite{Abe:1994mf}.  This property has a transparent explanation in perturbative QCD. At high  energy, the final states are dominated by two-jet events in which most of the energy is carried by particles moving along two different directions defining the jet axis. In this case, the sum in \re{EEC-def} receives a dominant contribution from energetic
particles % $a$ and $b$
 belonging either to the same jet (for $\chi\to 0$), or to the two different back-to-back jets (for $\chi\to \pi$). 

Being  an infrared safe observable, the EEC can be computed perturbatively in powers of the coupling constant. Neglecting hadronization corrections, we can take the sum in \re{EEC-def}  to run over the final states consisting of an arbitrary number of massless quarks and gluons. To find the corrections to \re{EEC-def} we have to compute the differential cross-section and, then, integrate it over the
available phase space with the energy weight factor. The calculation becomes rather cumbersome beyond the leading order in the coupling due  both to the intricate cancellation of infrared and collinear divergences at the intermediate steps and the necessity for summation over all final states.

The
leading-order (LO) correction to the EEC has been found four decades ago \cite{Basham:1978bw,Basham:1978zq}, the next-to-leading (NLO) correction was computed analytically only recently \cite{Dixon:2018qgp}.~\footnote{The next-to-next-to-leading correction to the EEC is known numerically~\cite{DelDuca:2016csb}.}
 It provides a reliable QCD prediction  for correlation angle $0< \chi <\pi$ away from the end-points.
In the end-point region,  perturbative corrections to the EEC are enhanced by powers of large logarithms. They originate from the emission of particles with momentum that is soft compared to the hard scale $Q$ and/or collinear to one of the energetic particles.  
 Such corrections have to be resummed to all orders. For  $\chi\to 0$ and $\chi\to \pi$, this can be done by employing the jet calculus \cite{Konishi:1978yx,Konishi:1978ax,Konishi:1979cb,Richards:1983sr} and the Sudakov resummation \cite{Collins:1981uk,Kodaira:1981nh,Ellis:1983fg,deFlorian:2004mp,Moult:2018jzp}, respectively. Both approaches rely on a careful separation of 
the contribution from particles with different momenta (hard, soft and collinear). The resulting expression for the EEC is given by 
a convolution of hard, jet and soft functions satisfying renormalization group evolution equations. This allows us to express the EEC in the end-point region in terms of a few functions of the coupling constant (see Eqs.~\re{resum0} and \re{resum} below). 

Another approach to computing the EEC was proposed in \cite{Belitsky:2013bja,Belitsky:2013xxa}. It relies on the description of the 
final states of $e^+e^-$ annihilation in terms of the   `energy flow operator' $\mathcal E(\vec n)$~\cite{Sveshnikov:1995vi,Korchemsky:1997sy,Korchemsky:1999kt,Belitsky:2001ij,Hofman:2008ar}. This operator
measures the energy flux per unit solid angle in a given
direction  $\vec n$ and it is built from the stress-energy tensor $T_{\mu\nu}(x)$ in a certain limit. It allows us to express the EEC in terms of the  Wightman (non-time-ordered) four-point
correlation function, schematically
\begin{align}\label{EEC-gen}
\text{EEC}(\chi) \sim \int d^4 x \e^{ix \cdot q} g^{\mu\nu}\vev{0|J_\mu(0)\mathcal E(\vec n_1)\mathcal E(\vec n_2)J_\nu(x) |0}\,,
\end{align}
where $q=(Q,\vec 0)$ is the total centre-of-mass momentum and the unit vectors $\vec n_1$ and $\vec n_2$ are separated by angle $\chi$. In this representation, the $U(1)$ current $J_\mu$ creates the final state out of vacuum~\footnote{In general, the Lorentz indices of the currents in \re{EEC-gen} should be contracted with the polarization vector of the incoming virtual photon. After averaging over the angular correlations between the final state and the incoming beams in \re{EEC-def}, the sum over the polarizations  gives $g^{\mu\nu}$.}  and the flow operators $\mathcal E(\vec n_i)$ describe the calorimeters.  

The representation \re{EEC-gen} holds in a generic four-dimensional Yang-Mills theory including QCD. In order to apply \re{EEC-gen}, we need the expression for the four-point correlation functions of the form $\vev{0|J_{\mu_1}(1) T_{\mu_2\nu_2} (2)T_{\mu_3\nu_3} (3)J_{\mu_4}(4) |0}$ where the stress-energy tensors come from the definition of the energy flow operators.~\footnote{In what follows we refer to $J_\mu$ and $T_{\mu\nu}$ as source and calorimeter operators, respectively.}  In general, these correlation functions have a complicated form. 

The situation simplifies if the underlying gauge theory enjoys conformal symmetry. In this case, 
 the correlation function depends on two cross-ratios and it can be described using the Mellin representation formalism \cite{Mack:2009mi}. The  energy-energy correlation \re{EEC-def} admits a very compact representation in terms of the corresponding Mellin ampltiude $M(j_1,j_2)$  \cite{Belitsky:2013bja,Belitsky:2013xxa}
\begin{align}\label{EEC-new}
{\rm EEC}(z)  = {1\over 2 (1-z)^3} \int {dj_1 dj_2 \over (2\pi i)^2}  
{ M(j_1,j_2)\Gamma(1-j_1-j_2)\over \Gamma(j_1+j_2)[\Gamma(1-j_1)\Gamma(1-j_2)]^2}
 \lr{z\over 1-z}^{-2-j_1-j_2},
\end{align}
where $z=\sin^2(\chi/2)$ depends on the correlation angle and the integration runs parallel to the imaginary axis. 
The relation \re{EEC-new} holds for $0<z<1$. To extend it to the end points, 
we have to add  to the right-hand side of \re{EEC-new} the Born level contribution $[\delta(z)+\delta(1-z)]/4$. Here the two terms correspond to $a=b$ and $a\neq b$ in the sum \re{EEC-def}.

At weak coupling, the representation \re{EEC-new} allows us to compute the EEC order-by-order in the coupling constant by replacing the  Mellin ampltiude $M(j_1,j_2)$ by its perturbative expansion.
The power of this approach was illustrated by obtaining for the first time an analytical NLO result \cite{Belitsky:2013ofa} and, very recently, NNLO \cite{Henn:2019gkr}  result for the EEC in maximally supersymmetric Yang-Mills theory
($\mathcal N=4$ SYM).  

In theories with broken conformal symmetry, the relation \re{EEC-new} holds up to corrections proportional to the beta function. 
The comparison of the NLO results for the EEC in $\mathcal N=4$ SYM \cite{Belitsky:2013ofa} and in QCD \cite{Dixon:2018qgp} shows that the two expressions have a very similar structure -- they involve the same type of (polylogarithmic) functions. This suggests that $\mathcal N=4$ SYM can be used to identify an appropriate basis of functions for the EEC in QCD beyond the leading order. Moreover, it was conjectured in \cite{Belitsky:2013ofa} that, in the back-to-back region, for $\chi\to\pi$, the EEC in $\mathcal N=4$ SYM describes 
the maximally transcendental part of the analogous QCD expression. 
   
In this paper, we apply \re{EEC-new} to study the properties of the EEC in the end-point regions, for $\chi\to 0$ and $\chi\to \pi$.  
We expect that, in both limits, the relation \re{EEC-new} should generate large logarithmically enhanced corrections. The resulting asymptotic expressions for the EEC should match the analogous expressions obtained using the QCD
resummation technique mentioned above. Thanks to (super) conformal symmetry, they take a particular simple form in $\mathcal N=4$ SYM:

For small  angle $\chi\to 0$, the jet calculus approach leads to~\footnote{Note that the jet calculus only describes the  leading logarithmic corrections of the form $a^{n+1}(\ln z)^n/z$. We show below that the relation \re{resum0} takes into account all logarithmically enhanced corrections.}  
\begin{align} \label{resum0}
{\rm EEC} (\chi) = \frac{h(a)}4\, z^{-1+\widehat \gamma_3(a)/2}\,,
\end{align} 
where $z=\sin^2(\chi/2)\to 0$ and the hard function $h(a)=a + O(a^2)$ depends on the coupling constant $a=g^2N/(4\pi^2)$.
It was argued in \cite{Hofman:2008ar} that $\widehat \gamma_3(a)$ coincides with the anomalous dimension of the $SU(4)$ singlet twist-two operator with Lorentz spin $3$.  
 
 For  large angle $\chi\to \pi$, in the back-to-back region, the Sudakov resummation yields
\begin{align}\label{resum}
{\rm EEC} (\chi) = \frac{H(a) }{8y} \int_0^\infty db \, b J_0(b)  
 \exp\left[ -\frac12 \Gamma_{\rm cusp}(a) \ln^2(b^2/(yb_0^2))-\Gamma(a) \ln (b^2/(yb_0^2)) \right],
\end{align}
where $y=1-z=\cos^2(\chi/2)$ vanishes as $\chi\to \pi$, $J_0(b)$ is a Bessel function and $b_0=2\e^{-\gamma_E}$ is a kinematical factor depending on the Euler constant. The relation \re{resum} depends on three functions of the coupling constant.
The functions $\Gamma_{\rm cusp}(a)$ and $\Gamma(a)$ govern the large spin asymptotics of the twist-two anomalous dimension (see \re{large-S} below),
they are known in planar $\mathcal N=4$ SYM for any coupling from integrability \cite{Beisert:2006ez,Freyhult:2007pz,Freyhult:2009my,Fioravanti:2009xt}.  The hard function $H(a)=1+O(a)$  does not have a simple interpretation, it is usually determined 
by matching the resummed expression \re{resum} to the result of a fixed order calculation. 

The relations \re{resum0} and \re{resum} were derived by analysing QCD evolution of particles in the final states of $e^+e^-$ annihilation. 
At the same time, the relation \re{EEC-new} does not have such an interpretation 
 -- it depends on the Mellin amplitude which defines the four-point correlation function in Euclidean space~\cite{Mack:2009mi}. 
 Given the complexity of the correlation functions and the (relative) simplicity of \re{resum0} and \re{resum},
 the question arises how could the representation \re{EEC-new} reproduce \re{resum0} and \re{resum} in the limit $z\to 0$ and $z\to 1$, respectively, or equivalently, which properties of the four-point correlation functions 
are probed in the two limits. 

In this paper, we answer this question by deriving \re{resum0} and \re{resum} from \re{EEC-new}. 
We show that the relation \re{resum0} follows from the resummation of the OPE of the two calorimeter operators.
In a similar manner, the relation \re{resum} follows from the asymptotic behavior of the four-point correlation function in the limit 
when the source operators are light-like separated from the calorimeter operators. 
As a byproduct of our analysis, we obtain closed expressions for the hard functions $h(a)$ and $H(a)$ in terms of the conformal data of twist-two operators. We verify that these expressions are in agreement with the NNLO result for the EEC in $\mathcal N=4$ SYM.
  
The paper is organized as follows. In section~\ref{sect2}, we analyze the relation \re{EEC-new} in the back-to-back limit $z\to 1$. We use the known result for the four-point correlation function in the light-like limit to reproduce the relation \re{resum}.
In section~\ref{sect3}, we apply the OPE to obtain a representation of the EEC at small angle as an infinite sum over the conformal partial waves
with even Lorentz spin. Evaluating this sum we arrive at \re{resum0}. In section~\ref{sect4}, we apply the approach of \cite{Lance} to derive 
sum rules for a regular part of the EEC and verify them at three-loop order.
Concluding remarks are presented in secton~\ref{sect5}. Appendix~\ref{sect:Hahn} contains some technical details.
  
\section{Energy correlations in the back-to-back region}\label{sect2} 

In this section, we analyze the relation \re{EEC-new} at large angle $\chi\to \pi$, or equivalently for $z\to 1$. We recall that the Mellin amplitude $M(j_1,j_2)$ defines the four-point correlation function of  
the form  $\vev{J_{\mu_1}(x_1) T_{\mu_2\nu_2}(x_2)T_{\mu_3\nu_3}(x_3)J_{\mu_4}(x_4) }$ where the $U(1)$ currents  excite the vacuum to produce the final states in $V \to a+ b+ X$ and the stress-energy tensors describe the calorimeters. Due to its nontrivial Lorentz structure, this correlation function has a rather complicated form in a generic four-dimensional conformal field theory. 

Significant simplification occurs in $\mathcal N=4$ SYM. The conserved currents in this theory belong to the same stress-energy supermultiplet whose lowest component is a scalar half-BPS operator of dimension two belonging to the  representation $\boldsymbol{20'}$  of
the $R$ symmetry $SU(4)$ group. It has the form $O_{IJ} = \tr(\phi_I \phi_J)-\delta_{IJ} \tr(\phi_K \phi_K)/6 $, where $\phi_I$ (with $I=1,\dots,6$) are real scalar
fields. 
As a consequence, the four-point correlation function of currents can be expressed in terms of the simplest correlation function of the form \cite{Belitsky:2014zha,Korchemsky:2015ssa}
\begin{align}\label{corr}
\vev{O(x_1) \widetilde O(x_2) \widetilde O(x_3) O(x_4)} = {\Phi(u,v)\over x_{12}^2 x_{13}^2 x_{24}^2 x_{34}^2}\,,
\end{align}
where $O(x)$ and $\widetilde O(x)$ are linear combinations of $O_{IJ}(x)$ and $x_{ij}^2=(x_i-x_j)^2$. Here the function $\Phi(u,v)$ depends on the cross-ratios
\begin{align}\label{uv}
u={x_{12}^2 x_{34}^2 \over x_{14}^2 x_{23}^2}\,,\qqqquad
v={x_{13}^2 x_{24}^2 \over x_{14}^2 x_{23}^2}\,,
\end{align}
and admits an expansion in powers of the coupling constant $a=g^2N/(4\pi^2)$. The Mellin amplitude entering \re{EEC-new} is defined as
\begin{align}\label{G-Mellin}
\Phi(u,v) = \int {dj_1 dj_2\over (2\pi i)^2} u^{j_1} v^{j_2} M(j_1,j_2)\,.
\end{align}
In general, $M(j_1,j_2)$ is a meromorphic function of the complex spins $j_1$ and $j_2$. The positions at the poles of $M(j_1,j_2)$ and the corresponding residues
encode the information about the behaviour of the correlation function \re{corr} in different OPE limits~\cite{Mack:2009mi}. 

\subsection{Moments of the energy-energy correlation}     

Let us show that for $\chi\to \pi$  the EEC  is controlled by the behaviour of the Mellin amplitude $ M(j_1,j_2)$ at small $j_1$ and $j_2$, or equivalently by asymptotics of $\Phi(u,v)$ for $u\to 0$ and $v\to 0$. For the correlation function \re{corr} this corresponds 
to the limit when the four operators become light-like separated in a sequential manner, $x_{12}^2,x_{13}^2,x_{24}^2,x_{34}^2 \to 0$.
     
\begin{figure}[t!]
\psfrag{x1}{$x_1$}\psfrag{x2}{$x_2$}\psfrag{x3}{$x_3$}\psfrag{x4}{$x_4$}\psfrag{q}{$V$}
\centerline{\includegraphics[width = 95mm]{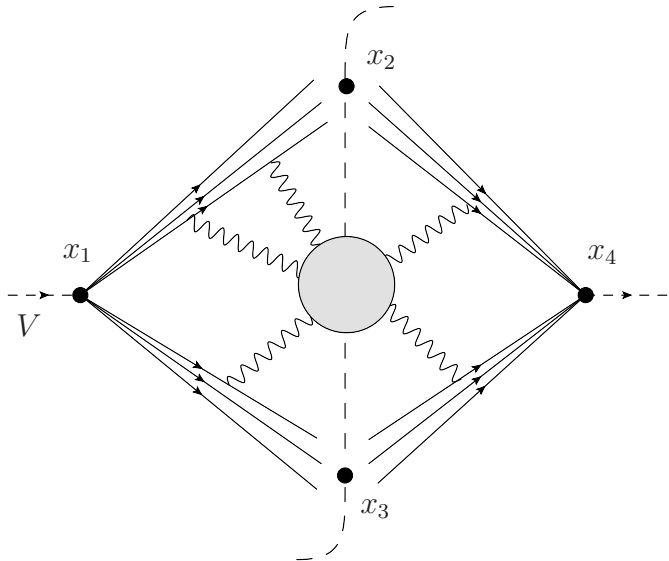}}
\caption{Unitary diagram describing a two-jet cross-section. The graphs to the left and to the right of the unitary cut (dashed line) describe the scattering amplitude $V\to \text{everything}$ and its complex conjugate counter-part. Solid and wavy lines denote collinear and soft particles, respectively. The grey blog describes the interaction between soft particles.
In the dual description, the `source operators' are located at the points $x_1$ and $x_4$ and the `calorimeter operators' are at the points $x_2$ and $x_3$.}
\label{2jets}
\end{figure}

As was mentioned in the Introduction,  for $\chi\to \pi$, the energy-energy correlation receives the leading contribution from the two-jet events. 
In this case, the final state consists of two nearly back-to-back jets accompanied by a soft radiation (see Fig,~\ref{2jets}). 
Since the contribution of soft particles with energy $E_a$ to \re{EEC-def} is suppressed by the factor of $E_a/Q\ll 1$, the sum in \re{EEC-def} effectively runs over pairs of particles belonging to two different jets. 
The correlation angle $\chi$ fixes the angular separation between their spacial momenta.

The two-jet cross section admits the following dual description in terms of the correlation function \re{corr} (see Refs.~\cite{Erdogan:2014gha,Erdogan:2017gyf} for a description of scattering amplitudes in coordinate space). The source operator  $O(x_1)$  creates  fast particles which propagate in the direction of the two calorimeter operators $\widetilde O(x_3)$ and $\widetilde O(x_4)$ and, then, get absorbed by the sink operator $O(x_4)$ (see Fig.~\ref{2jets}). In the high-energy limit, the particles propagate close to the light-cone and, therefore, the source/sink operators become light-like separated from the calorimeter operators. Notice that the particles carry the color charge of the adjoint representation of the $SU(N)$. As a consequence, they source a radiation that back reacts on them. It consists of soft particles as well as collinear particles whose momenta are aligned with the momenta of the parent particles. In the next subsection, we use this picture to compute the energy correlations in the back-to-back region.
 
It proves convenient to introduce
the moments of the energy-energy correlations  
\begin{align}\label{mom}
\widetilde {\rm EEC} (N) 
{}& = \int_0^1 d z \, z^{N-1} \, {\rm EEC} (z) 
\\\label{mom-zero}
{}&= \int {dj_1 dj_2 \over (2\pi i)^2}  
{ \Gamma(1-j_1-j_2)\Gamma(N-2-j_1-j_2)\over 2\,[\Gamma(1-j_1)\Gamma(1-j_2)]^2\Gamma(N-2)} M(j_1,j_2)
\,,
\end{align}
where $z=\sin^2(\chi/2)$ and in the second relation we replaced  ${\rm EEC} (\chi)$ with its expression \re{EEC-new}. 
Notice that due to the factor of $\Gamma(N-2)$ in the denominator, $\widetilde {\rm EEC} (N)$ vanishes for $N=1$ and $N=2$. 
We are going to exploit this property in section~\ref{sect:sumrules}. 
 
Because at large $N$  the integral in \re{mom} receives a dominant contribution from the end-point region $z\sim 1$, the EEC for $\chi\to\pi$ can be found from the large $N$ 
behaviour of the moments. Replacing the ratio of $\Gamma-$function in \re{mom-zero} by their leading asymptotics at large $N$ we find  for $N\gg 1$
\begin{align}\label{repr1}
\widetilde {\rm EEC}(N) = \int {dj_1 dj_2 \over (2\pi i)^2}  
{ \Gamma(1-j_1-j_2)\over 2\,[\Gamma(1-j_1)\Gamma(1-j_2)]^2} M(j_1,j_2) N^{-j_1-j_2}\,.
\end{align}
At large $N$ the leading contribution to the integral comes from $j_{1,2}=O(1/\ln N)$. This allows us to replace the ratio of $\Gamma-$functions on the right-hand side of \re{repr1} by its expansion at small $j_i$ 
\begin{align}\label{Gam-dec}
{\Gamma(1-j_1-j_2)\over [\Gamma(1-j_1)\Gamma(1-j_2)]^2} = \e^{-\gamma_E ( j_1+j_2)}\left[1-{\zeta_2\over  2}(j_1-j_2)^2
%-{\zeta_3\over 3}(j_1^3-3 j_1^2 j_2-3j_1j_2^2+j_2^3)
 + \dots \right]
\end{align}
and expand the moments \re{repr1} in powers of $1/\ln N$.  

The leading term of the expansion comes from the first term inside the brackets in \re{Gam-dec}
\begin{align}\label{LO}
\widetilde {\rm EEC}(N)   &=  \frac12 \int {dj_1 dj_2 \over (2\pi i)^2} M(j_1,j_2) \bar N^{-j_1-j_2}  +\dots \,,
= \frac12\Phi (u,v)\Big|_{u=v=1/\bar N} +\dots
\end{align}
where $\bar N=N \e^{\gamma_E}$  and the dots denote subleading corrections suppressed by powers of $1/\ln N$. Here in the second
relation we applied \re{G-Mellin} and identified the Mellin integral as the function $\Phi(u,v)$ evaluated for the special values of the cross-ratios $u=v=1/\bar N$ with $\bar N\gg 1$. It is straightforward to include the subleading corrections to $\widetilde {\rm EEC}(N)$. They  come from the remaining terms inside the brackets of \re{Gam-dec} given by homogenous polynomials in $j_1$ and $j_2$. 
Notice that powers of $j_1$ and $j_2$ can be generated by applying logarithmic derivatives with respect to $u$ and $v$ to both sides of \re{G-Mellin}. Therefore, in order to produce subleading corrections to $\widetilde {\rm EEC}(N)$, it suffices to replace $j_1\to u\partial_u$ and $j_2\to v\partial_v$ in \re{Gam-dec},  apply the resulting differential operator to the correlation function $\Phi (u,v)$ and replace the cross-ratios with their values  $u=v=1/\bar N$. In the next subsection, we derive an expression for $\widetilde {\rm EEC}(N) $ that takes into account all subleading corrections.

We conclude that, in agreement with our expectations, the energy-energy correlation  in the back-to-back region is related to the asymptotics of the correlation function \re{corr} in the light-cone limit  $u=v=1/\bar N$  with $\bar N =N \e^{\gamma_E}\to\infty$. 

\subsection{Correlation function in the light-cone limit}\label{sect:H}
   
The properties of the correlation functions \re{corr} in the light-cone limit $x_{12}^2,x_{23}^2,x_{34}^2,x_{41}^2 \to 0$ have been studied in Refs.~\cite{Alday:2010zy,Alday:2013cwa}. In this subsection, we summarize the finding of these papers and, then, apply them to compute \re{repr1} at large $N$.

In the light-like limit, the four operators in \re{corr} approach the vertices of a light-like rectangle. In the Born approximation, for zero coupling, the correlation function  \re{corr} is given by the product of four singular scalar propagators connecting the points $x_i$. 
%  leading to  $\Phi(u,v)=1$. 
It describes the propagation of a free scalar particle along the edges of the light-like rectangle. Turning on the interaction, we have to take into account the back reaction of the radiation that  the particle creates. This gives the following factorized expression for $\Phi(u,v)$ in the limit $u,v\to 0$
\begin{align}\label{G-fact}
\Phi(u,v) = \frac12 H\, S(u,v)\, J(u,v)\,,
\end{align}
where each factor on the right-hand side describes the contribution of particles with particular momenta (hard, soft and collinear) and the factor of $1/2$ is introduced for convenience. The hard function only depends on the coupling constant. It describes creation/annihilation of fast particles at points $x_i$.

The interaction of the fast particles with soft radiation gives rise to the soft function. It can be expressed in terms of an (appropriately regularized) 
rectangular light-like Wilson loop and is given by the Sudakov form factor 
\begin{align}\label{S-2}
S(u,v) =  \exp\left[-\frac12 \Gamma_{\rm cusp}(a) \ln (u c_0) \ln(v c_0)+\frac12 \Gamma(a) \ln (u c_0)+\frac12 \Gamma(a) \ln (v c_0)  \right]\,,
\end{align}
where $c_0=(b_0/2)^2=\e^{-2\gamma_E}$ is a kinematical factor depending on the Euler's constant. Although this relation holds in a conformal field theory, it can be generalized to theories with a nonvanishing beta-function~\cite{Mueller:1979ih,Sen:1981sd,Korchemsky:1988hd,Collins:1989bt}. 

The expression \re{S-2} depends on two functions of the coupling constant known as the cusp and collinear anomalous dimensions. The same functions
control the asymptotic behaviour of the anomalous dimension of twist-two operators with  large spin~\cite{Korchemsky:1988si}
\begin{align}\label{large-S}
\gamma_S(a) = 2 \Gamma_{\rm cusp}(a) \lr{\ln S+ \gamma_E} +\Gamma(a) + O(1/S)\,,
\end{align} 
and their OPE coefficients~\cite{Eden:2012rr,Alday:2013cwa} \footnote{For a twist-two operator $O_S$ these coefficients are schematically  defined as 
$C_S\sim \vev{O\widetilde O O_{S}}\vev{O_S\widetilde O O}/\vev{O_S O_S}$. }
\begin{align}\label{C-large}
C_S(a)/C_S(0) = F^2(a)  \e^{-\Gamma(a) (\ln S+\gamma_E)- \gamma_S(a) \gamma_E} 2^{-\gamma_S(a)}\Gamma^2\left(1-\frac12\gamma_S(a)\right) + O(1/S)\,,
\end{align}
where $F(a)$ is independent on the spin and $C_S(0)$ is the coefficient function at  zero coupling (see \re{data0} below).
We would like to stress that these relations hold for $\gamma_S(a)/2<1$. For larger values of $\gamma_S(a)$, the twist-two operators collide with the twist-four operators and the relations \re{large-S} and \re{C-large} are modified \cite{Korchemsky:2015cyx}. 

The relation \re{G-fact} can be derived by applying the OPE to the correlation function \re{corr} in the channels $x_{12}^2\to 0$ and $x_{13}^2\to 0$. In the light-cone limit, for $u\to 0$ and $v\to 0$,  the dominant contribution in both channels comes from twist-two operators with large spin $S$. It can be found explicitly by taking into account \re{large-S} and \re{C-large}.
In this way, we  reproduce \re{G-fact} and \re{S-2} and, in addition, obtain a prediction for the
hard and collinear functions in terms of the conformal data (the OPE coefficients $C_S$ and anomalous dimensions $\gamma_S$) of the twist-two operators.  

The resulting expression for $\Phi(u,v)$ takes a remarkably simple form~\cite{Alday:2010zy,Alday:2013cwa}
\begin{align} \label{G-int}
\Phi(u,v) {}& = \frac12F^2(a) \int_0^\infty {dy_1\over y_1}  \int_0^\infty {dy_2\over y_2} \,  S\left({u\over y_1},{v\over y_2}\right)  f(y_1)f(y_2) \,,
\end{align}
where $S(u,v)$ is the soft function \re{S-2} and the hard function $H=F^2(a)$ coincides with the constant, $S-$independent part of the OPE coefficients \re{C-large}.
The relation \re{G-int} can be interpreted as a double OPE expansion of the
correlation function. Namely, the integral over $y_1$ comes from the sum over the twist-two operators with large spins $S_1=y_1/\sqrt{u}$ propagating in the OPE channel $x_{12}^2\to 0$. The function $f(y_1)$ describes the large spin limit of the corresponding  conformal block  (see \re{block} below)
\begin{align} 
f(y) =  2 y K_0(2\sqrt{y})\,,
\end{align}
where $K_0$ is the modified Bessel function of the second kind.
It satisfies the following relation
\begin{align}\label{f-M}
\int_0^\infty {dy\over y} y^{-j} f(y)= \left[\Gamma(1-j)\right]^2\,.
\end{align}
In the same manner, the integral over $y_2$ in \re{G-int} comes from the sum over the twist-two operators with large spins $S_2=y_2/\sqrt{v}$ in the OPE channel $x_{23}^2\to 0$.
  
The relations \re{G-int} and \re{S-2} describe the asymptotic behaviour of the four-point correlation function in the light-cone limit $u,v\to 0$. They depend on
three functions of the coupling constant, $\Gamma_{\rm cusp}(a)$, $\Gamma(a)$ and $F(a)$, which can be extracted from the conformal
data \re{large-S} and \re{C-large} of twist-two operators with large spin $S$. In $\mathcal N=4$ SYM, the first two functions are known to all loops  in the planar limit 
from integrability \cite{Beisert:2006ez,Freyhult:2007pz,Freyhult:2009my,Fioravanti:2009xt}~\footnote{Nonplanar corrections appear starting from four loops.}   whereas the last one is currently known to three loops~\cite{Eden:2012rr,Alday:2013cwa} . 
  
\subsection{Comparison with the Sudakov resummation}    
    
We can now combine together \re{G-Mellin}, \re{repr1} and \re{G-int} and derive the Sudakov resummation formula \re{resum}
for the energy-energy correlation in the back-to-back region. 

To this end, it is convenient to Mellin transform the soft function \re{S-2}
\begin{align}\label{S-M}
S(u,v) = \int {dj_1 dj_2\over (2\pi i)^2} u^{j_1} v^{j_2} \widetilde S(j_1,j_2)\,.
\end{align}
Substituting this relation into \re{G-int} and taking into account \re{f-M} we get
\begin{align}
\Phi(u,v) = \frac{H(a)}2 \int {dj_1 dj_2\over (2\pi i)^2 } u^{j_1} v^{j_2} \widetilde S(j_1,j_2) \left[\Gamma (1-j_1)\Gamma(1-j_2)\right]^2\,.
\end{align}
Comparing this relation with \re{G-Mellin} we conclude that the Mellin amplitudes of the Sudakov form factor and the correlation function are related as $M(j_1,j_2)= H(a)\widetilde S (j_1,j_2)\left[\Gamma (1-j_1)\Gamma(1-j_2)\right]^2\!\!/2 $.
We then obtain from \re{S-M} 
\begin{align} \label{HS}
H(a)\, S(u,v) = \int {dj_1 dj_2\over (2\pi i)^2 } {2 M(j_1,j_2) \, u^{j_1} v^{j_2}\over \left[\Gamma (1-j_1)\Gamma(1-j_2)\right]^2}\,.
\end{align}
We would like to emphasize that this relation holds for $u,v\to 0$. 
\footnote{It is interesting to note that the expression on the left-hand side of \re{HS} differs from \re{G-fact} by the jet factor. At the same time,
the  Mellin integrals in \re{HS} and \re{G-Mellin} are very similar to each other and the only difference between them is the product of $\Gamma-$functions in the denominator of \re{HS}.}

We are now ready to compute the moments of the energy-energy correlation \re{repr1} from \re{HS}. To match the Mellin integral on the right-hand side of \re{repr1}, it is sufficient to put $u=v=1/N$  and apply the operator $\Gamma(1+N\partial_N)/4$ to both sides of \re{HS} 
\begin{align}\notag\label{mom-fin}
\widetilde {\rm EEC}(N) {}&= \frac14\Gamma(1+N\partial_N) H(a)\, S(u,v)\Big|_{u=v=1/N} 
\\
{}& =\frac14 H(a) \int_0^\infty dx\, \e^{-x} S\left(1/(N x)\right). %\Big|_{u=v=1/(N x)}
\end{align}
Here in the second relation we replaced the $\Gamma-$function by its integral representation and took into account that the 
operator $x^{N\partial_N}$ generates dilatations $N\to x N$. We also introduced the notation for the soft function \re{S-2} with  coinciding arguments
\begin{align}\label{Su}
S(u) \equiv S(u,u) =  \exp\left[-\frac12 \Gamma_{\rm cusp}(a) \ln^2 (u c_0)+ \Gamma(a) \ln (u c_0)  \right]\,,
\end{align}
where $c_0=\e^{-2\gamma_E}$. Substituting this relation into \re{mom-fin} we obtain an expression for $\widetilde {\rm EEC}(N) $ that takes into account all the logarithmically enhanced contibutions to the moments of the energy-energy correlation at large $N$.

Let us now invert the moments \re{mom-fin} and reproduce \re{resum}.  
We recall that at large $N$ the integral in \re{mom} receives a dominant contribution from the end-point region $z\sim 1$. This allows us to replace $z^N=(1-y)^N \approx \exp(-y N)$
and extend the integration over $y$ to infinity
\begin{align}\label{mom0}
\widetilde {\rm EEC} (N) = \int_0^\infty d y \, \e^{-yN} {\rm EEC} \,.
\end{align}
We can bring \re{mom-fin} to this form by changing the integration variable in \re{mom-fin} as  $x= b^2/(4N)$ and applying the identity
\begin{align}
  {1\over N}\e^{-b^2/(4N)} = \int_0^\infty dy\, J_0(b y^{1/2}) \e^{-y N}\,.
\end{align}
This leads to  
\begin{align}
\widetilde {\rm EEC}(N) =  {H(a)\over 8}\int_0^\infty dy\, \e^{-yN}  \int_0^\infty db\, b J_0(by^{1/2}) S(4/b^2)\,.
\end{align}
Matching this relation into \re{mom0} and replacing $S(4/b^2)$ with \re{Su}, we arrive at an expression for the EEC that coincides with 
 \re{resum}. 

\subsection{Checks}

The energy-energy correlation has been computed in $\mathcal N=4$ SYM at the NLO order in \cite{Belitsky:2013ofa} and very recently  at the NNLO order in~\cite{Henn:2019gkr}. 
Expanding  \re{resum} to order $O(a^3)$, we should be able to reproduce all the logarithmically enhanced terms in the NNLO expression for the EEC 
in the end-point region  $\chi\to \pi$. 

The explicit expressions for the anomalous dimensions are~\cite{Kotikov:2004er,Beisert:2006ez,Freyhult:2007pz,Freyhult:2009my,Fioravanti:2009xt}
\begin{align}\notag\label{3loops}
& \Gamma_{\rm cusp}(a) =  a-\frac{\pi ^2 }{12}a^2+\frac{11 \pi ^4 }{720}a^3- \left(\frac{\zeta
   _3^2}{8}+\frac{73 \pi ^6}{20160}\right)a^4+O (a^5 )\,,
\\[2mm] 
& \Gamma(a) =-\frac{3  \zeta _3}{2}a^2+ \left(\frac{\pi ^2 \zeta _3}{12}+\frac{5 \zeta
   _5}{2}\right)a^3-\left( \frac{7\pi ^4 \zeta _3}{480}  +\frac{5 \pi ^2 \zeta
   _5}{48}+\frac{175 \zeta _7}{32}\right)a^4+O (a^5 )\,,
\end{align}
where $a=g^2 N/(4\pi^2)$ and the expansion coefficients depend on zeta values $\zeta_n$ with different weights $n$. The expansion coefficients  of $\Gamma_{\rm cusp}$ and $\Gamma$ have uniform weights $2\ell-2$ and $2\ell-1$, respectively, at order $O(a^\ell)$.

As explained in Sect.~\ref{sect:H}, the hard function $H=F^2(a)$ can be extracted from the OPE coefficients of twist-two operators $C_S$ in the  large spin limit. The OPE coefficients $C_S$ for arbitrary $S$ have been computed to three loops in \cite{Eden:2012rr} and their asymptotics for $S\gg 1$ was found in \cite{Alday:2013cwa}. Using  the results  of these papers we get
\begin{align}\label{hard}
H(a) {}& =1-\frac{\pi ^2 }{6}a+\frac{\pi ^4 }{18}a^2-\left(\frac{17 \zeta
   _3^2}{12}+\frac{197 \pi ^6}{10080}\right)a^3 +O (a^4 )\,,
\end{align}
where the expansion coefficients have uniform weight $2\ell$ at order $O(a^\ell)$. 

Substituting  \re{3loops} and \re{hard} into \re{resum} and expanding the EEC in powers of the coupling constant we get 
\begin{align}\notag\label{EEC-dec}
\text{EEC}   = {1\over 4y}  {}& \bigg\{ a L   + a^2 \lr{ -\frac{L^3}{2}-\frac{\pi ^2 L}{4}+\frac{\zeta _3}{2}} 
  \\ \notag  {}&
  +a^3 \left( \frac{L^5}{8}+\frac{\pi ^2 L^3}{6}-\frac{11 \zeta _3
   L^2}{4}+\frac{61 \pi ^4 L}{720}-\frac{\pi ^2 \zeta
   _3}{3}-\frac{7 \zeta _5}{2}\right)
\\\notag
& \qquad + a^4 \bigg[ -\frac{L^7}{48}-\frac{5 \pi ^2
   L^5}{96}+\frac{95 \zeta _3 L^4}{48} 
  -\frac{29 \pi ^4 L^3}{480} +\left(\frac{67 \pi ^2
   \zeta _3}{48}+\frac{69 \zeta _5}{4}\right) L^2
\\
{}&   
 \qqqquad  -\left(\frac{97 \zeta
   _3^2}{8}+\frac{367 \pi ^6}{12096}\right) L+\frac{187 \pi ^4 \zeta _3}{1440}+\frac{95 \pi ^2
   \zeta _5}{48}+\frac{785 \zeta _7}{32} \bigg]+O(a^5)\bigg\}\,,
\end{align}
where $L=\ln(1/y)$ and $y=\cos^2(\chi/2)$ vanishes for $\chi\to\pi$. The first three terms of the expansion \re{EEC-dec} are in agreement with the results of Refs.~\cite{Belitsky:2013ofa,Henn:2019gkr}, the last term yields  a prediction for the N${}^3$LO correction to the EEC in 
$\mathcal N=4$ SYM. 

Assigning  weight $1$  to $L=\ln(1/y)$ we observe that the expansion coefficients in \re{EEC-dec} have uniform weight $2\ell-1$ at order $O(a^\ell)$. Notice that this property does not hold for a generic angle $\chi$, e.g. the NLO result for the energy-energy correlation is given by a linear combination of polylogarithms of mixed weight.   
In the back-to-back region, the contribution of  functions of  lower weight is suppressed and the uniform weight property is restored.

As  mentioned in the Introduction,  in the back-to-back region, the EEC in $\mathcal N=4$ SYM should describe 
the maximally transcendental part of the analogous QCD expression (upon identification of the color factors $C_F\to N$)  \cite{Belitsky:2013ofa}. Assuming that this conjecture holds at higher loops, we can apply  \re{EEC-dec} to predict 
the maximal weight part of the EEC in QCD.

 \section{Energy correlations at small angle}\label{sect3}

At small angle $\chi$, the EEC measures the correlation between the fast particles within the same jet. 
Invoking the dual description of the energy correlations in terms of the correlation function \re{corr},  we expect that
the calorimeter operators $\widetilde O(x_2)$ and $\widetilde O(x_3)$ should be located at short distances.

We can arrive at the same conclusion by examining \re{EEC-new} for $z=\sin^2(\chi/2)\to 0$
\begin{align} \label{EEC-small}
{\rm EEC}(\chi)  = \int {dj_1 dj_2 \over (2\pi i)^2}  
{ \Gamma(1-j_1-j_2)\over 2\Gamma(j_1+j_2)[\Gamma(1-j_1)\Gamma(1-j_2)]^2}M(j_1,j_2)
 z^{-j_1-j_2-2}\,,
\end{align}
where we replaced $z/(1-z)= z + O(z^2)$ and neglected subleading corrections.
Ignoring the ratio of $\Gamma-$functions in \re{EEC-small} for the sake of the argument, we can apply  \re{G-Mellin} and 
express the resulting Mellin integral in terms of the function $\Phi(u,v)$ evaluated at large values of the cross-ratios $u=v=1/z$ as $z\to 0$.
As follows from \re{uv}, the limit of large $u$ and $v$ is realized for $x_{23}^2\to 0$.

\subsection{Conformal partial wave expansion}

Discussing the properties of the correlation functions \re{corr} for $x_{23}^2\to 0$,
it is convenient to redefine the cross-ratios \re{uv}
\begin{align}\label{uv1}
u'={x_{23}^2 x_{14}^2 \over x_{13}^2 x_{24}^2}=1/v\,,\qqqquad
v'={x_{12}^2 x_{34}^2 \over x_{13}^2 x_{24}^2} =u/v\,,
\end{align}
so that $u'\to 0$ and $v'\to 1$  in the limit $x_2\to x_3$. Then, the correlation function \re{corr} and  \re{G-Mellin} takes the form
\begin{align}\label{G-Mellin1}
\vev{O(x_1) \widetilde O(x_2) \widetilde O(x_3) O(x_4)} = {1\over (x_{23}^2 x_{41}^2)^2} 
\int {dj_1 dj_2\over (2\pi i)^2} (v')^{j_1-1} (u')^{-j_1-j_2+2} M(j_1,j_2)\,.
\end{align}
For $u'\to z$ and $v'\to 1$, the Mellin integral on the right-hand side differs from the analogous integral in \re{EEC-small} %and \re{G-Mellin1} only differ
 by the product of $\Gamma-$functions.
To elucidate their role, we expand the correlation function on the left-hand side of \re{G-Mellin1} over the conformal partial waves in the channel $x_{23}^2\to 0$.
 
The OPE expansion of the correlation function \re{G-Mellin1} has been studied in Refs.~\cite{Arutyunov:2001mh,Dolan:2001tt,Eden:2012fe}. It was shown there that the 
leading contribution to  $\widetilde O(x_2) \widetilde O(x_3)$ for $x_{23}^2\to 0$ comes from the twist-six operator 
carrying the $R-$charge of the representation $\boldsymbol{105}$ of the $SU(4)$ group. 
\footnote{Strictly speaking, the leading contribution to the OPE comes from a double-trace protected operator $\widetilde O(x_2) \widetilde O(x_2)$ of twist four. Its contribution to the four-point correlation function \re{cpwe} does not depend on the coupling constant and, as a consequence, it does not affect  the EEC. }
This operator belongs to the twist-two supermultiplet
and, as a consequence, its conformal data (the scaling dimension and the OPE coefficient) are related to those of the twist-two operator,
$\Delta_{\boldsymbol{105}}(S) = 6+ S + \gamma_{S+2}$ and $C_{\boldsymbol{105}}(S) = C_{S+2}$. 
Its contribution to the four-point function is given by
\begin{align} \label{cpwe}
& \vev{O(x_1) \widetilde O(x_2) \widetilde O(x_3) O(x_4)} = {1\over (x_{23}^2 x_{41}^2)^2} \sum_{S/2\in \mathbb Z_+}
C_{S+2}(a) (u')^{3+\gamma_{S+2}(a)/2} g_S(v')+\dots\,,
\end{align}
where the sum runs over even nonnegative spins $S$ and the dots denote subleading terms suppressed by powers of $u'$.
Here $\gamma_S$ and $C_S$ are the conformal data 
%of the   operator $\tr(ZD_+^S Z)+\dots $ (with $Z=\phi_1+i\phi_2$) defining the lowest weight state 
of the twist-two $\mathcal N=4$ supermultiplet
 and $g_S(v')$ is the collinear conformal block
\begin{align}\label{block}
g_S(v')=(1-v')^S {}_2 F_1 \lr{3+S+\ft12 \gamma_{S+2}(a), 3+S+\ft12 \gamma_{S+2}(a); 6+2S+ \gamma_{S+2}(a)|1-v'}\,.
\end{align}
Comparing \re{G-Mellin1} and \re{cpwe}, we find that the Mellin amplitude satisfies %the following relation
\begin{align}\label{eq-M}
\int {dj_1 dj_2\over (2\pi i)^2} (v')^{j_1-1} z^{-j_1-j_2-2} M(j_1,j_2) = \sum_{S/2\in \mathbb Z_+}
C_{S+2}(a) z^{-1+\gamma_{S+2}(a)/2} g_S(v')+O(z^0)\,,
\end{align}
where we replaced $u'=z$. We can use this relation to evaluate the Mellin integral in \re{EEC-small}. 
 
For $v'=1$ the Mellin integrals on the left-hand side of \re{EEC-small} and \re{eq-M} only differ by the product of $\Gamma-$functions. 
Since the integrand in \re{eq-M} has a power-like dependence on $v'$ and $z$, this product can be generated by acting on \re{eq-M} 
with the following operator
\begin{align}\label{op}
{\Gamma(3+ z\partial_z)\over 2\Gamma(-2-z\partial_z)[\Gamma(-v'\partial_{v'})\Gamma(4+ z\partial_z +v'\partial_{v'})]^2}\,.
\end{align}
 To apply this operator to the right-hand side of \re{eq-M}, it is convenient to replace the conformal block \re{block} with its Mellin
 representation~\cite{Costa:2012cb}
 \begin{align}\label{block-M}
 g_S(v') =  
  &  \frac{\Gamma (2S+\tau)}{ \left[\Gamma
   \left( {\tau}/{2}\right) 
   \Gamma \left(S+ {\tau}/{2}\right)\right]^2}
\int{dj \over 2\pi i} (v')^{j}  \left[\Gamma(-j)  \Gamma(j+\tau/2)\right]^2 
  \, {}_3 F_2 \left({-S,S+\tau-1, -j\atop \tau/2,\tau/2} \Big| 1\right),
\end{align}
where  $\tau=6+\gamma_{S+2}$ is the twist of the exchanged operator. Here the integration contour runs parallel to the imaginary axis and separates the poles of $\Gamma(-j)$ and $\Gamma(j+\tau/2)$. In what follows we choose $j=-\tau/4 + i x$ with $-\infty<x<\infty$.

Substituting \re{block-M} into \re{eq-M} we find that the action of the  
operator \re{op} on the right-hand side of \re{eq-M} amounts to introducing a factor of $\Gamma(\tau/2-1)/(2\left[\Gamma(-j)  \Gamma(j+\tau/2)\right]^2\Gamma(2-\tau/2))$ inside the Mellin integral in \re{block-M}. In this way, we obtain from \re{EEC-small} and \re{eq-M}
\begin{align}\label{EEC-sum}
\text{EEC}=2\int_{-\infty}^\infty{dx \over 2\pi}  \sum_{S/2\in \mathbb Z_+}z^{\tau/2-4}
  A_{S,\tau}
 P_{S,\tau}(x)\,,
\end{align} 
where $\tau=6+\gamma_{S+2}$ and the notation was introduced for
\begin{align}\notag\label{Hahn}
{}& A_{S,\tau} =
 \frac{ \Gamma (2S+\tau)\Gamma(\tau/2-1)}{4\left[\Gamma
 ( \tau/2) 
   \Gamma \left(S+ {\tau}/{2}\right)\right]^2\Gamma(2-\tau/2)}C_{S+2}(a)\,,
\\[2mm]
{}& P_{S,\tau}(x)=  \, {}_3 F_2 \left({-S,S+\tau-1, \tau/4 - i x\atop \tau/2,\tau/2} \Big| 1\right)\,.
\end{align}   
For integer nonnegative $S$ the function $ P_{S,\tau}(x)$ is a polynomial in $x$ of degree $S$ with definite parity
\begin{align}\label{par}
P_{S,\tau}(-x) = (-1)^S  P_{S,\tau}(x)\,.
\end{align}
Up to a normalization factor, it coincides with the continuous Hahn polynomial  \cite{Koekoek}.

Each term in the sum \re{EEC-sum}  is polynomial in $x$ and, therefore, it yields a divergent contribution upon integration over $x$. 
For the integral in \re{EEC-sum} to be well-defined, the sum over the spins should be regular on the real
$x-$axis and decrease sufficiently fast at infinity.
It is convenient to introduce the auxiliary function
\begin{align}\label{f-aux}
f(x) = \sum_{S=0}^\infty z^{\tau/2-4}
 A_{S,\tau} 
 P_{S,\tau}(x)\,,
\end{align}
where the sum runs over all nonnegative spins. Using \re{par} we can rewrite \re{EEC-sum} as
\begin{align}\label{EEC-con}
\text{EEC} = \int_{-\infty}^\infty{dx \over 2\pi} \left[ f(x) + f(-x)\right]\,.
\end{align}
In the next subsection, we show that $f(x)$ is a meromorphic function with poles located in the lower half-plane and a simple pole at infinity
\begin{align}\label{f-infty}
f(x) = {i f_{\infty}\over x} + O(1/x^2)\,.
\end{align}
The sum of the functions $f(x) + f(-x)$ has poles on both sides of the real axis and decreases at infinity as $O(1/x^2)$. This allows us to close the integration contour in \re{EEC-con} into the lower half-plane and compute the integral by the residues at all the poles of $f(x)$ except the one at infinity. Since the sum of all residues of $f(x)$ equals zero, the integral is then given by the residue at infinity
\begin{align}\label{EEC-inf}
\text{EEC} = f_\infty\,.
\end{align}
To find $f_\infty$, we have to examine the asymptotics of \re{f-aux} at  large $x$ and match it into \re{f-infty}. This is done in Sect.~\ref{sect:jet}.

\subsection{Warm up example}
   
Let us examine \re{f-aux} to the lowest order in the coupling.  Replacing in \re{Hahn}
\begin{align}\label{tau}
\tau=6 + \gamma_{S+2} = 6 + a \gamma^{(0)}_{S+2} + O(a^2)
\end{align}
and taking into account that $1/\Gamma(2-\tau/2)=O(a)$, we find  
\begin{align}\label{A-1}
A_{S,\tau} =  a  
 \frac{ \Gamma (2S+6)}{32 
   \Gamma^2 \left(S+ 3\right)}C^{(0)}_{S+2}\gamma^{(0)}_{S+2} + O(a^2)\,.
\end{align}
Here $C^{(0)}_{S+2}$ and $\gamma^{(0)}_{S+2}$ are the leading order expressions for 
 the conformal data of the twist-two operators with spin $S+2$ \cite{Dolan:2001tt}  
\begin{align}\label{data0}
C^{(0)}_{S+2} = {\Gamma^2(S+3)\over \Gamma(2S+5)} \,,\qqqquad
\gamma^{(0)}_{S+2} = 2\left[\psi(S+3)-\psi(1)\right]\,,
\end{align}
where $\psi(x)= (\ln \Gamma(x))'$ is the Euler $\psi-$function.

Substituting \re{tau} and \re{A-1} into \re{f-aux} we get
\begin{align} \label{mech1}
f(x) {}& = {a\over 16z} \sum_{S=0}^\infty (5S+5) \left[\psi(S+3)-\psi(1)\right] P_{S,6}(x) + O(a^2)
\\\label{mech}
{}& =  {a\over 4z} \left[ {i\over x+\frac{i}2} - {{1\over 2(x+\frac{i}2)^2}}\right]+ O(a^2)\,,
\end{align}
where the sum in the first relation can be evaluated using the properties of the continuous Hahn polynomial (see Appendix~\ref{sect:Hahn} for details). 
  
The relation \re{mech} illustrates the main feature of the function $f(x)$ mentioned in the previous subsection. Namely, each term in the sum   \re{mech1} is polynomial in $x$  but the function has a (double) pole at $x=-i/2$. It arises from the large spin contribution. For $S\gg 1$ we have  $P_{S,6}(x) \sim S^{-3+2ix}$ (see \re{P-as}), so that the sum \re{mech1} scales as
\begin{align}\label{f-poles}
f(x)\sim \int_{S_0}^\infty dS\, S^{-2+2ix} \ln S \sim 1/(1-2ix)^2\,,
\end{align}
where $S_0\gg 1$.

At large $x$ we compare \re{mech} with \re{f-infty} and \re{EEC-inf} to get
\begin{align}
\text{EEC} = {a\over 4z} + O(a^2)\,.
\end{align}
This relation is in agreement with the known LO result for the energy-energy correlation \cite{Engelund:2012re}
\begin{align}\label{EEC-LO}
\text{EEC} = - a{\ln(1-z) \over 4 z^2(1-z)} + O(a^2) 
\end{align}
in the end-point region $z=\sin^2(\chi/2)\to 0$. 

\subsection{Comparison with the jet calculus}\label{sect:jet}

Following the analysis in the previous subsection, we can understand analytical properties of the function $f(x)$ defined in \re{f-aux} by examining the large spin contribution. For $S\gg 1$ the anomalous dimensions and the OPE coefficients are given by \re{large-S} and \re{C-large}, respectively, whereas the continuous Hahn polynomial scales as (see \re{P-as})
\begin{align}\label{P-as1}
P_{S,\tau} (x) \sim S^{-\tau/2+2ix} \,,
\end{align}
where $\tau=6+\gamma_{S+2}$. Substituting these relations into \re{f-aux} we find that at large $S$ the function $f(x)$  is given by an integral similar to \re{f-poles}. At weak coupling, perturbative corrections to $C_{S+2}$ and $\tau$ modify the integrand of \re{f-poles} by terms of the form $a^n \ln^k S$. Such terms generate higher order poles $a^n/(x+i/2)^{2+k}$ but they do not alter analytical properties of $f(x)$.

Let us now find the leading asymptotic behaviour of the function \re{f-aux} at infinity. For $x\to\infty$ we find from \re{Hahn} and \re{par}
\begin{align}\notag\label{P-large}
P_{S,\tau}(x) {}& %=(-1)^S P_{S,\tau}(-x) 
=(-1)^S\sum_{k\ge 0} {(-S)_k (S+\tau-1)_k\over [(\tau/2)_k]^2} {(ix)^k\over k!}
\\
{}& = {(-1)^S\Gamma^2(\tau/2) \over\Gamma(-S)\Gamma(S+\tau-1) }\int {dj\over 2\pi i} {\Gamma(-j)\Gamma(-S+j)\Gamma(S+\tau-1+j)\over   \Gamma^2(\tau/2+j)}(-ix)^j\,,
\end{align}
where $(a)_k=\Gamma(a+k)/\Gamma(a)$ is the Pochhammer symbol. Here in the first relation we used a series representation of the hypergeometric function and replaced $(\tau/4+ix)_k\sim (ix)^k$ at large $x$. In the second relation, we converted the sum into 
a Mellin-Barnes integral. 

Substituting \re{P-large} into \re{f-aux}  we can employ the Sommerfeld-Watson transformation 
to rewrite the sum over spins as
\begin{align}\label{f-double}
f(x)  = \int {d S \,dj \over (2\pi i)^2} z^{\tau/2-4} A_{S,\tau}  {\Gamma(S+1)\Gamma^2(\tau/2)\Gamma (-j)\Gamma(j-S)\Gamma(j+S+\tau-1)\over \Gamma(S+\tau-1)\Gamma^2(\tau/2+j)}(-ix)^j\,,
\end{align}
where the integration over $j$ runs parallel to the imaginary axis and separates the poles generated by the functions
$\Gamma(j+\dots)$ and $\Gamma(-j+\dots)$. The integration contour over $S$ is defined in a similar manner.

To find the leading asymptotics of \re{f-double} at large $x$ we move the integration contour over $j$ to the left and pick up the residue at the right-most pole. This pole is located at $j=S$ and its contribution to \re{f-double} is
\begin{align}\label{f-single}
f(x)  = \int {d S  \over 2\pi i } z^{\tau/2-4} A_{S,\tau}  {\Gamma(S+1)\Gamma^2(\tau/2)\Gamma (-S) \Gamma(2S+\tau-1)\over \Gamma(S+\tau-1)\Gamma^2(\tau/2+S)}(-ix)^S+ \dots\,,
\end{align}
where the dots denote subleading terms suppressed by powers of $1/x$.
Then, for the same reason as before, we move the integration contour over $S$ to the left and pick up the residue at the leading pole $S=-1$.
  In this way, we arrive at
\begin{align}\label{f-res}
f(x) = - {1\over  ix} \left(  z^{\tau/2-4} A_{S=-1,\tau}   { (\tau/2 -1)^2\over  \tau-3 } \right) + O(1/x^2)\,,
\end{align}
where the expression inside the brackets is evaluated for the unphysical value of spin $S=-1$ and $\tau=6+\gamma_{1}(a)$.

The relation \re{f-res} has the expected form \re{f-infty}. We apply \re{EEC-inf} together with \re{Hahn} and obtain the following 
result for the leading asymptotic behaviour of the  EEC in the end-point region $z=\sin^2(\chi/2)\to 0$
\begin{align}\label{EEC-pre1}
\text{EEC} 
 {}& = z^{\tau/2-4}
  \frac{C_{1}(a) \Gamma (\tau-3)}{4 \Gamma^3 \left(\tau/2-1\right) \Gamma(2-\tau/2)}   
 \\ 
 \label{EEC-pre}
{}& = z^{-1+\gamma_1(a)/2}   \frac{ C_{1}(a) \,\Gamma (3+\gamma_1(a))}{4 \Gamma^3 \left(2+\gamma_1(a)/2\right) \Gamma(-1-\gamma_1(a)/2)}\,,
\end{align}
where in the second relation we replaced $\tau=6+\gamma_1(a)$.
Here $\gamma_1(a)$ and $C_1(a)$ are, respectively,  the anomalous dimension and the OPE coefficient of the twist-two operator analytically continued to Lorentz spin $S=1$. 
 
The relations  \re{EEC-pre1} and \re{EEC-pre}  agree with the analogous expressions previously derived in \cite{Sasha}  using a different approach based on the OPE expansion of light-ray operators~\cite{Kravchuk:2018htv,Kologlu:2019bco}.

Comparing \re{EEC-pre} with the analogous relation \re{resum0} predicted by the jet calculus, we
notice that they involve the twist-two anomalous dimensions $\gamma_1$ and $\widehat\gamma_3$, evaluated for different values of  Lorentz spin, $S=1$ and $S=3$, respectively. The reason for this is that the anomalous dimensions $\gamma_1$ and $\widehat\gamma_3$ correspond of two different operators -- the lowest weight of  the twist-two $\mathcal N=4$ supermultiplet and the $SU(4)$ singlet operator of the form $\tr(\phi_I D_+^S \phi_I) + \dots$, respectively. The latter operator belongs to the same supermultiplet and its anomalous dimension is related to that of the former as $\widehat\gamma_S=\gamma_{S-2}$ (see \cite{Dolan:2001tt,Belitsky:2003sh}). Thus, the relations \re{resum0} and \re{EEC-pre} have the same $z-$dependence of the EEC. The relation \re{EEC-pre} also allows us to predict the hard function entering \re{resum0} 
\begin{align}\label{h}
h(a) %= \frac{ C_{1}(a) \,\Gamma (3+\gamma_1(a))}{\Gamma^3 \left(2+\gamma_1(a)/2\right) \Gamma(-1-\gamma_1(a)/2)}
=\frac{C_{1}(a)\, \gamma_1(a) \, \Gamma (2+\gamma_1(a))}{\Gamma^2
   \left(1+ \frac12{\gamma_1(a) } \right) \Gamma \left(2+\frac12 {\gamma_1(a) } \right)\Gamma \left(1- \frac12{\gamma_1(a) }\right) }
\,.
\end{align} 
In the next subsection, we test this relation using available results for the EEC in the small angle limit.
 
 The relation \re{EEC-pre} describes the leading behaviour of the EEC at small $z$. The expression on the right-hand side of \re{EEC-pre} arises from the contribution to the OPE \re{cpwe} of the conformal primary operator with the minimal twist $\tau=6+O(a)$. Taking into account the contribution to \re{cpwe} of the operators with higher twist $\tau=2(n+3)+O(a)$ (with $\ge 1$), we can generate subleading 
 $O(z^{n-1})$ corrections to \re{EEC-pre}. They take the form \re{EEC-pre1} with $C_1$ being the OPE coefficient for the operator of twist $\tau$. Another source of subleading corrections to the EEC has a kinematical origin and has to do with the fact that the relation \re{EEC-small} was obtained from \re{EEC-new} by replacing $z/(1-z)=z+ O(z^2)$ and $1/(1-z)^3=1+O(z)$. To incorporate these corrections, it is sufficient to  substitute $z\to z/(1-z)$ on the right-hand side of \re{EEC-pre1} and insert the additional factor of $1/(1-z)^{3}$ (see Eq.~\re{EEC-new}).
   
\subsection{Checks}

Expanding \re{h} in powers of $\gamma_1(a)$ we get
\begin{align}\label{h-pred}
h(a) = C_{1}(a) \left( \gamma_1 +\frac{\gamma_1 ^2}{2}-\frac{\gamma_1 ^3}{4}+ \left(\frac{1}{8}-\frac{\zeta_3}{4}\right) \gamma_1 ^4 +O (\gamma_1 ^5 ) \right)\,.
\end{align}
The weak coupling expansion of $C_1(a)$ and $\gamma_1(a)$ starts at order $O(a^0)$ and $O(a)$, respectively.
To compute $h(a)$ at order $O(a^\ell)$ we need the expressions for $C_1(a)$ and $\gamma_1(a)$ at $(\ell-1)$ and $\ell$ loops, respectively.

To find $C_1(a)$ and $\gamma_1(a)$ we have to analytically continue the expressions for the OPE coefficients and anomalous dimensions of the twist-two operators from even spins  to $S=1$. 
To the leading order in the coupling, this can be easily done using \re{data0}.   
Starting from the next-to-leading order, the expressions for $C_S(a)$ and $\gamma_S(a)$ involve alternating nested harmonic sums $S_{\pm a,\pm b,\dots} $. These sums depend on the sign factors $(-1)^S$ and their continuation from even and odd spins gives rise to two different functions $\widebar S^+_{\pm a,\pm b,\dots} $ and $\widebar S^-_{\pm a,\pm b,\dots} $, respectively~\cite{Kotikov:2005gr}. In order to compute $C_1(a)$ and $\gamma_1(a)$, 
we only need the former function.

The twist-two anomalous dimension $\gamma_S$ can be computed for even $S$ and any coupling in planar $\mathcal N=4$ SYM from integrability. For our purpose, we use the analytical expression for $\gamma_S$  to order $O(a^4)$ from Refs.~\cite{Kotikov:2007cy,Bajnok:2008qj}. Following ~\cite{Kotikov:2005gr}, we can analytically continue it to $S=1$ to get
the four-loop result for $\gamma_1$~\footnote{I would like to thank Lance Dixon for his help in computing $\gamma_1$ and $C_1$.} 
 \begin{align}\label{g1}\notag
 \gamma_1(a)  {}& =
2 a+a^2 \left(-\zeta _3+\frac{\pi ^2}{3}-4\right)
\\ {}&\notag
+a^3 \left(3 \zeta
   _5+\frac{\pi ^4}{120}-3 \zeta _3-\frac{4 \pi ^2}{3}+16\right)
\\   \notag
{}&\qquad   +a^4\bigg(-\frac{69 \zeta _7}{8}+\frac{\pi ^2 \zeta _5}{6}-\frac{\pi ^4 \zeta _3}{144} +\frac{9 \zeta
   _3^2}{2}-\frac{107 \pi
   ^6}{15120}
\\ {}&    \qqqquad 
  \  +16 \zeta _5 -\frac{13 \pi ^2 \zeta _3}{6} -\frac{23 \pi ^4}{360} +14 \zeta
   _3+8 \pi ^2-80\bigg)  +O (a^5) \,,
\end{align}
where $a=g^2N/(4\pi^2)$. This relation is in agreement with the findings of \cite{Lance}.

The twist-two OPE coefficients $C_S(a)$ are currently known to order $O(a^3)$~\cite{Eden:2012rr}. \footnote{The OPE coefficients are also known at order $O(a^4)$ for spin $S=2,4,6,8$ \cite{Eden:2016aqo}. These results can be used to find a numerical interpolation for $S=1$.}
Their analytical continuation to $S=1$ yields the three-loop result for $C_1$
\begin{align}\label{C1}\notag
 C_1(a) {}& = \frac12- a+a^2 \left(\frac{1}{4}\zeta_3+\frac{11 \pi ^4}{720}-\frac{ \pi
   ^2}{3}+6\right)
 \\  
  {}& +a^3 \left(\frac{3 \zeta _3^2}{8}-\frac{109 \pi ^6}{15120}+6 \zeta _5-\frac{7 \pi ^2 \zeta _3}{12} -\frac{\pi ^4}{48}+3 \zeta
   _3+3 \pi
   ^2-40\right)+O(a^4)\,. 
\end{align} 
Finally, we substitute \re{g1} and \re{C1} into \re{h-pred} to get the hard function
\begin{align}\label{h-fun}\notag
   h(a)  {}&= a+a^2 \left(-\frac{\zeta_3}{2}+\frac{\pi ^2}{6}-3\right)
   \\ {}& \notag
   +a^3 \left(\frac{3
   \zeta_5}{2}+\frac{5 \pi ^4}{144}-\zeta_3 -\frac{4 \pi ^2}{3}+17\right)
 \\  \notag
   {}&\qquad +a^4 \left( -\frac{69 \zeta _7}{16}+\frac{\pi ^2 \zeta
   _5}{12}-\frac{3 \pi ^4 \zeta _3}{160} + 3 \zeta
   _3^2 -\frac{389 \pi ^6}{30240} \right.
 \\  {}&\qqqquad \left.
   \ +20 \zeta _5-2 \pi ^2 \zeta _3    -\frac{3 \pi
   ^4}{16}+10 \zeta _3 +\frac{65 \pi ^2}{6}-111\right) +O (a^5 )\,.
\end{align}
The first three terms of the expansion are in agreement with the NNLO result for the energy-energy correlation in
$\mathcal N=4$ SYM \cite{Belitsky:2013ofa,Henn:2019gkr},
 the $O(a^4)$ term is a prediction.
 
\section{Sum rules for the energy correlations} \label{sect:sumrules}\label{sect4}

As follows from the definition \re{EEC-def}, the EEC satisfies the normalization condition
\begin{align}\label{sum1}
\int_0^1 dz\, \text{EEC}(z) = \frac12 \int_{-1}^1 d(\cos\chi) \, \text{EEC}(\chi)  = {(\sum_a E_a)^2\over 2Q^2} =\frac12\,,
\end{align}
where $\sum_a E_a=Q$ is the total  energy.
It was shown in \cite{Lance} that this relation leads to interesting consistency conditions for the singular behaviour of the EEC in the limits $z\to 0$ and $z\to 1$. In this section, we apply the ideas of \cite{Lance} to derive nontrivial relations for
a `regular' part of the EEC that is obtained by subtracting from $\text{EEC}(z)$ the terms singular at the end-points.
The analogous relations were also derived in \cite{Sasha} using the light-ray OPE.

Notice that the integral on the left-hand side of \re{sum1} coincides with the moments \re{mom}  for $N=1$. There is however an important difference -- in distinction from \re{sum1}, the relation \re{mom-zero} does not take into account the Born contribution to the EEC. It is given by $(\delta(z)+\delta(1-z))/4$ and automatically satisfies \re{sum1}. 
Therefore, subtracting the Born contribution on both sides of \re{sum1} we deduce that the moments \re{mom} have to vanish for $N=1$. Indeed, it is easy to see  from \re{mom-zero} that $\widetilde{\text{EEC}}(1)=0$. 
As mentioned above, the relation \re{mom-zero}  implies that $\widetilde{\text{EEC}}(2)=0$. Adding the Born level contribution, we arrive at the sum rule for the second moment of the EEC
\begin{align}\label{sum2}
\int_0^1 dz z\,  \text{EEC}(z) %= \frac14 \int_{-1}^1 d(\cos\chi) (1-\cos\chi)\, \text{EEC}(\chi) 
=\frac14\,,
\end{align}
where $z=(1-\cos\chi)/2$. This relation has a simple interpretation. Using the definition \re{EEC-def}, the integral in \re{sum2} can be evaluated as
\begin{align}
\sum_{a,b} {E_a E_b\over 4Q^2} (1-\cos\theta_{ab}) = \sum_{a,b} {(p_a p_b)\over 4Q^2} = \frac14\,,
\end{align}
where $\sum_a p_a=(Q,\vec 0)$ is the total momentum.

It is convenient to split the EEC into the sum of singular and regular terms
 \begin{align}
\text{EEC}(z) = \text{EEC}_0(z) + \text{EEC}_{\rm reg}(z) + \text{EEC}_1(z)\,,
\end{align}
where $\text{EEC}_0(z)$ and $\text{EEC}_1(z)$ describe the singular behaviour for $z\to 0$ and $z\to 1$, respectively.
They  have the following general form
\begin{align}\notag\label{Vs}
& \text{EEC}_0(z) = \delta(z) V_0(a) + \left[{ \varphi_0(z) \over z} \right]_+,
 \\
 & \text{EEC}_1(z) = \delta(1-z) V_1(a) + \left[{\varphi_1(1-z)\over 1-z}   \right]_+,
\end{align}
where $\varphi_0(z) $ and $\varphi_1(1-z)$ are given by the sum of logarithmically enhanced terms of the form
$a^{k+1} (\ln z)^n$ and $a^{k+1} (\ln (1-z))^n$, respectively.
Here the notation was introduced for the `+' distribution
\begin{align}
\int_0^1 dz \,  w(z) \left[{ \varphi_0(z)\over z}  \right]_+ = \int_0^1 dz \, \left[w(z)-w(1) \right]{ \varphi_0(z) \over z}\,,
\end{align}
where $w(z)$ is a test function. The distribution $\left[{\varphi_1(1-z)/(1-z)}   \right]_+$ is defined in the same manner.
As compared to \re{resum0} and \re{resum}, the relations \re{Vs} contain the additional terms proportional to $\delta(z)$ and $\delta(1-z)$. They describe the contribution from the special configurations when, respectively, the same particle goes through the two detectors and the final state consists of two back-to-back particles. To the lowest order in the coupling, we have $V_0=\frac14+O(a)$ and $V_1=\frac14+O(a)$. 

By construction, the function  $\text{EEC}_{\rm reg}(z)$ does not contain singular $O(1/z)$ and $O(1/(1-z))$ terms for $z\to 0$ and $z\to 1$, respectively.
In what follows we refer to it as a regular part of the EEC. To the leading order in the coupling, 
we find from \re{EEC-LO}
\begin{align}\label{EEC-reg}
\text{EEC}_{\rm reg}(z) = {a\over 4}\left[- {\ln(1-z) \over z^2(1-z)} -{1\over z}+{\ln(1-z)\over 1-z}\right]+ O(a^2) \,.
\end{align}
Substituting \re{Vs} into the relations \re{sum1} and \re{sum2},  we get the sum rules for the first two moments of the regular part of the EEC
\begin{align}\notag\label{sum-reg}
 & \int_0^1 dz\, \text{EEC}_{\rm reg} =\frac12 -V_0-V_1\,,
\\
 & \int_0^1 dz z\,\text{EEC}_{\rm reg} = \frac14 -  V_1  -\int_0^1 dz\,\varphi_0(z) +  \int_0^1 dy\,\varphi_1(y) \,,
\end{align}
where $y=1-z$.
At zero coupling, the functions $\varphi_0(z) $, $\varphi_1(z) $ and $\text{EEC}_{\rm reg}(z)$ vanish
leading to $V_0(0)=V_1(0)=\frac14$, in agreement with the Born level contribution.

To find the functions entering the right-hand side of \re{sum-reg}, we have to match the ansatz \re{Vs} to the leading singular behavior of the EEC at the end-points, Eqs.~\re{resum0} and \re{resum}.
For $z\to 0$ we have $\text{EEC}\sim h(a) \, z^{-1+\gamma_1/2}/4$.  Applying the identity
\begin{align}
{1\over z^{1-\epsilon}} = \frac1{\epsilon}\delta(z) + \left[1\over z^{1-\epsilon} \right]_+\,,
\end{align}
we arrive at the first relation in \re{Vs} upon identification
\begin{align}\label{set1} \notag
& \varphi_0(z)= \frac14 h(a) \, z^{\gamma_1(a)/2}\,,
\\
& V_0 = \int_0^1 {dz\over z} \varphi_0(z) ={h(a) \over 2\gamma_1(a)} 
\,.
\end{align}
For $z\to 1$, the functions in \re{Vs} are known from the Sudakov resummation. Namely,  $\varphi_1(y)/y$ coincides with \re{resum} for $y=1-z$ and $V_1(a)$ is proportional to the hard function 
\begin{align}\label{set2}\notag
& \varphi_1(y) =  {H(a)\over 8} \int_0^\infty db\, b J_0(b) S(4y/b^2)\,,
\\
& V_1= \int_0^1  {dy\over y} \, \varphi_1(y)  = {H(a)\over4}  \int_0^\infty db\,  J_1(b) S(4/b^2)\,,
\end{align}
where the Sudakov form factor $S(4y/b^2)$ is given by \re{Su}. 

Substituting the relations \re{set1} and \re{set2} into the sum rules \re{sum-reg} we find after some algebra
\begin{align} \label{sum-rules}\notag
 \int_0^1 dz\, \text{EEC}_{\rm reg}(z) {}& =\frac12- \frac14 H(a) \xi_0(a) -  {h(a)\over 2\gamma_1(a)}\,,
\\[2mm]   
 \int_0^1 dz  z\,\text{EEC}_{\rm reg}(z) {}&=\frac14-\frac14 H(a) (\xi_0(a)+\xi_1(a)) - {h(a)\over 4 +2\gamma_1(a)} \,.
\end{align} 
Here the notation was introduced for integrals involving the Bessel functions and the Sudakov form factor
\begin{align} \notag
 \xi_0(a) {}& = \int_0^\infty {dx\over \sqrt{x}}\, \e^{-\gamma_E} J_1(2\sqrt{x} \e^{-\gamma_E})   \e^{-\frac12\Gamma_{\rm cusp}(a) \ln^2 x- \Gamma(a) \ln x}
= 1- \frac{\zeta_3^2}3 a^3 + O(a^4)\,,
\\ 
  \xi_1(a) {}&= \int_0^\infty dx \, J_2(2\sqrt{x}\e^{-\gamma_E}) {\partial\over \partial x} 
\e^{-\frac12\Gamma_{\rm cusp}(a) \ln^2 x- \Gamma(a) \ln x} 
\\\notag
{}& =-a+a^2 \left(3+\frac{\pi ^2}{12}-\frac{\zeta_3}{2}\right)-a^3 \left(15+\frac{\pi ^2}{2}-\frac{11 \zeta
 _3}{2}+\frac{11
   \pi ^4}{720}-\frac{\pi ^2 \zeta_3}{4}-\frac{7 \zeta_5}{2}  \right)+O (a^4 ) \,,
\end{align} 
where we replaced the cusp and collinear anomalous dimensions with their expressions \re{3loops}.

We recall that $\text{EEC}_{\rm reg}(z)$ is obtained from the EEC by subtracting all terms that are singular for $z\to 0$ and $z\to 1$. 
This function describes the EEC away from the end-points and it is not captured neither by the Sudakov resummation, nor by the jet calculus. 
It is therefore remarkable that its first two moments \re{sum-rules}  can be expressed in terms of a few functions of the coupling constant controlling the singular behavior of the EEC in the end-point region. 
  
The expressions on the right-hand side of \re{sum-rules} depend on the hard functions, $h(a)$ and $H(a)$, and the anomalous dimension $\gamma_1(a)$.
Using their explicit expressions 
 \re{hard}, \re{g1} and \re{h-fun}, we obtain the sum rules at order $O(a^3)$
\begin{align}\notag\label{sum1-3loops}
 & \int_0^1 dz\, \text{EEC}_{\rm reg}(z) 
   =  \left(\frac{1}{4}+\frac{\pi ^2}{24}\right) a+\left(-\frac{7}{4}+\frac{\pi
   ^2}{8}-\frac{31 \pi ^4}{1440}\right) a^2
\\
 &\qqqquad   +\left(\frac{49}{4}-\pi ^2-\frac{5 \zeta_3}{4}+\frac{\pi
   ^4}{576}+\frac{7 \pi ^2 \zeta
  _3}{24}-\frac{27 \zeta_5}{8}+\frac{1027 \pi ^6}{120960}+\frac{\zeta_3^2}{4}\right)
   a^3+O\left(a^4\right)\,,
\\[2mm] \notag\label{sum2-3loops}
 & \int_0^1 dz z \, \text{EEC}_{\rm reg}(z) 
  = \frac{\pi ^2}{24}a+\left(\frac{1}{4}-\frac{5 \pi
   ^2}{48}+\frac{\zeta_3}{4}-\frac{\pi ^4}{72}\right) a^2 
\\
 & \qqqquad  +\left(-2+\frac{2 \pi
   ^2}{3}-\frac{11 \zeta_3}{8}+\frac{\pi ^4}{80}-\frac{\pi ^2
   \zeta_3}{12}-\frac{5 \zeta_5}{4}+\frac{197 \pi ^6}{40320}+\frac{7 \zeta_3^2}{16}\right)
   a^3+O\left(a^4\right)\,.
\end{align}
The sum rule \re{sum1-3loops} agrees at order $O(a^2)$ with the results of \cite{Lance}.
Using \re{EEC-reg}  it is straightforward to verify the relations \re{sum1-3loops} and \re{sum2-3loops} at order $O(a)$. At order $O(a^2)$, the function $\text{EEC}_{\rm reg}(z)$ can be expressed in terms of classical polylogarthms  \cite{Belitsky:2013ofa}. 
Evaluating its first two moments, we reproduced the $O(a^2)$ terms on the right-hand side of \re{sum1-3loops} and \re{sum2-3loops}.
At order $O(a^3)$, the function $\text{EEC}_{\rm reg}(z)$ has a complicated form -- it is given by a sum of harmonic polylogarithms plus
a two-fold finite (elliptic) integral \cite{Henn:2019gkr}. This makes an analytical calculation of its moments problematic. Evaluating them numerically, we reproduced the $O(a^3)$ correction to  \re{sum1-3loops} and \re{sum2-3loops} within one percent accuracy. The 
relations  \re{sum1-3loops} and \re{sum2-3loops} were also verified numerically to high accuracy in \cite{Sasha}.
This provides a stringent test of the approach described above.

\section{Concluding remarks}\label{sect5}

In this paper, we have studied the energy-energy correlation in the end-point region.
Our starting point was the representation \re{EEC-new} of the EEC in terms of the Mellin amplitude of the  four-point correlation function of the conserved currents. We used the properties of the correlation functions to find this observable in $\mathcal N=4$ SYM in the limit of small and large angles. 
In both cases, we obtained a concise representation of the EEC in terms of the conformal data of the twist-two operators.
We verified that the obtained expressions are in a perfect agreement both with the available results of the explicit calculation of the EEC at weak coupling and with the
analogous expressions obtained using QCD resummation techniques. 
We would like to emphasize that the above analysis was made under a tacit assumption that the spectrum of scaling dimensions  is sparse, allowing us to neglect a high-twist contribution. This assumption is justified at weak coupling but it does not hold at strong coupling due to a level-crossing phenomenon \cite{Korchemsky:2015cyx}. Indeed, the final states in $\mathcal N=4$ SYM do not 
contain jets at strong coupling 
and the EEC becomes a flat function of the angle \cite{Hofman:2008ar}.

We argued that, in the back-to-back region, the EEC is governed by the asymptotic behaviour of the four-point correlation function
in the limit when the operators are light-like separated in a sequential manner.  In this limit, the correlation function can be 
expressed in terms of a light-like rectangular Wilson loop and a hard function. This property is rather general -- it follows from the resummation of the leading, twist-two contribution to the OPE in different channels and it does not rely on the conformal symmetry. We demonstrated that
the resulting expression for the EEC in the back-to-back region coincides with the one predicted by the Sudakov resummation. 
We found that the hard function in $\mathcal N=4$ SYM is closely related to the asymptotics of the OPE coefficients of the twist-two operators with large spin and used this relation to predict the maximally transcendental part of the hard function in QCD.

Analyzing the EEC in the back-to-back limit $z\to 1$, we took into account the logarithmically enhanced terms and ignored the corrections
suppressed by powers of $(1-z)$. Such corrections are known to have a more complicated form. In the present approach, the power suppressed corrections can be systematically taken into account by including the high-twist contribution to the OPE in the light-like limit. 

In the small angle limit, we computed the EEC in $\mathcal N=4$ SYM by expanding the correlation function over the conformal partial waves in the channel corresponding to the product of calorimeter operators. This leads to a representation for the EEC as an infinite sum over the spins of the exchanged states. We used the properties of the conformal blocks to show that the sum is given by the contribution of the state with unphysical value of the spin $S=-1$. 

According to \re{EEC-pre}, the $z-$dependence of the EEC in the small angle limit is governed by the twist-two anomalous dimension $\gamma_1(a)$. The same anomalous dimension controls the scale dependence %of the second moment 
of the parton distribution function
in deep inelastic scattering and it is known as the space-like anomalous dimension.
In the leading logarithmic (LL) approximation, with $\gamma_1(a)$ replaced by its one-loop expression $\gamma_1(a)=2a+ O(a^2)$, the relation \re{EEC-pre} agrees with the prediction of the jet calculus.  Notice that the factor of $z^{\gamma_1(a)/2}$ in \re{EEC-pre} arises  in this approach
from the scale dependence of the parton fragmentation function and, therefore, one would naively expect that $\gamma_1(a)$ should have the meaning of a time-like anomalous dimension.  
At one loop, the time-like ($\gamma^{(T)}_N$) and space-like  ($\gamma^{(S)}_N$) anomalous dimensions coincide. They differ  starting from two loops and are related to each other as \cite{Basso:2006nk}
\begin{align}\label{S-T}
\gamma_N^{(S)}=\gamma^{(T)}_{N+\gamma_N^{(S)}} \,.
\end{align}
 It is therefore counter-intuitive that \re{EEC-pre} depends on 
the space-like anomalous dimension. % $\gamma_1(a)$. 

The resolution of the puzzle goes as follows~\cite{Simon}. 
The EEC at small angles $\chi$ measures the correlation between the two partons within the jet with the relative transverse momentum
$p_\perp^2=Q^2\chi^2\sim Q^2 z$. Let $D_N(Q^2/p_\perp^2)$ be the probability in the moment space to find the pair of partons with
the tranverse momentum $p_\perp^2$ and virtuality up to $Q^2$. It satisfies the QCD evolution equation \cite{Dokshitzer:2005bf,Marchesini:2006ax}
\begin{align}\label{EQ}
 Q {\partial  \over \partial  Q } D_N (Q^2/p_\perp^2) = \int_0^1 dx\, x^{N-1}  P_T(x) D_N (x^2 Q^2/p_\perp^2) \,,
\end{align}
where $P_T(x)$ is a time-like splitting function describing the fragmentation of a parton with a fraction $x$ of the longitudinal momentum. 
The EEC is related to $D_N(Q^2/p_\perp^2)$ for $N=3$ and $p_\perp^2=z Q^2$. 
The additional factor of $x^2$ in the argument of $ D_N (x^2 Q^2/p_\perp^2)$ on the right-hand side of \re{EQ} takes into account the evolution of the virtuality of the parton during the fragmentation process. 
In the LL approximation, we can neglect this factor and solve the evolution equation as $D_N(Q^2/p_\perp^2) \sim (Q^2/p_\perp^2)^{-\gamma^{(T)}_N/2}$, where $\gamma_N^{(T)} = -\int_0^1 dx\, x^{N-1} P_T(x)$ is the time-like anomalous dimension. Going beyond this approximation, we seek for a solution to \re{EQ} in the form $D_N(Q^2/p_\perp^2) \sim (Q^2/p_\perp^2)^{-\gamma_N/2}$. Its substitution into \re{EQ} yields
\begin{align}
 \gamma_N = -\int_0^1 dx\, x^{N-1 + \gamma_N}  P_T(x) =   \gamma_{N+\gamma_N}^{(T)}\,.
\end{align}
Comparing this relation with \re{S-T} we conclude that $\gamma_N=\gamma_N^{(S)}$. Thus,  in agreement with \cite{Hofman:2008ar}, the EEC has a power law behaviour at small $z$ with the power given by the space-like anomalous dimension.
The same result was obtained in \cite{Lance} from a factorization formula for the EEC in the small angle limit.

The relation \re{EEC-gen} establishes the connection between the EEC and the four-point correlation function of the conserved currents. It can be thought as a generalization of the optical theorem for the energy-energy correlation. 
We would like to emphasize that this relation does not rely on conformal symmetry and it also valid in QCD. This leads to the following important consequences.  
The very fact that the EEC is related to the Euclidean correlation function (upon a nontrivial analytical  continuation to its Wightman
counterpart  in Minkowski space-time) implies that its asymptotic behavior in the end-point region is controlled by the  space-time quantities.  This property is not specific feature of $\mathcal N=4$ SYM and it should also hold in QCD.

We remind that the representation analogous to \re{EEC-gen}
 holds for the total cross-section $\sigma_{\rm tot}$ of the process $V\to\text{everything}$. By virtue of the optical theorem, it is given 
by the  imaginary part of the two-point correlation function of the $U(1)$ currents. This representation allows us not only 
to compute efficiently perturbative corrections to $\sigma_{\rm tot}$ in QCD \cite{Baikov:2012zn} but also to describe the leading nonperturbative corrections to $\sigma_{\rm tot}$ in terms a few nonperturbative scales (QCD vacuum condensates) \cite{Shifman:1978bx,Mueller:1984vh}. In the similar manner, the representation 
\re{EEC-gen} can be used to study the nonperturbative corrections to the energy-energy correlation in QCD.
These interesting problems require further investigation.
 
\section*{Acknowledgements}
 
I am grateful to Simon Caron-Huot, Lance Dixon, Emery Sokatchev and Sasha Zhiboedov for interesting discussions. I am indebted to the authors of  Refs.~\cite{Sasha,Lance} for comparing their results for the EEC at small angle with mine and for sharing a draft of their papers before 
publication.
 This work was supported by the French National Agency for Research grant ANR-17-CE31-0001-01.

\appendix

\section{Continuous Hahn polynomial}\label{sect:Hahn}

For a nonnegative integer $S$, the function $P_{S,\tau}$ defined in \re{Hahn} coincides (up to a normalization factor) with the continuous Hahn polynomial \cite{Koekoek}.
It satisfies the finite-difference equation
\begin{align}\label{recrel}
2 i x (2 S+\tau -1)P_{S,\tau}(x) =  (S+\tau -1) (S+\tau/2 )P_{S+1,\tau}(x)  -S (S+\tau/2 -1)P_{S-1,\tau}(x) \,,
\end{align}
where $\tau$ is an arbitrary parameter. In what follows we consider $\tau$ to be independent on $S$.

At large $S$, the polynomial  scales as $P_{S,\tau}(x) \sim S^{\alpha}$. Substituting this ansatz into \re{recrel} and 
matching the leading large $S$ asymptotics we find $\alpha=-\tau/2+2ix$, in agreement with \re{P-as1}. We can obtain the same result by 
using the Mellin-Barnes representation for the hypergeometric function in \re{Hahn}
\begin{align}
P_{S,\tau}(x) = \int{dj\over 2\pi i} {\Gamma(-j)  \Gamma(\tau/4-ix +j) \Gamma^2(\tau/2)  \over \Gamma(\tau/4-ix) \Gamma^2(\tau/2+j) }  S^{2j} + \dots\,,
\end{align}
where we replaced the Pochhammer symbols $(-S)_j (S+\tau-1)_j$ by their leading large $S$ asymptotics. 
Here the integration contour runs parallel to the imaginary axis and separates the poles of the two $\Gamma-$functions in the numerator.
Moving the integration contour to the left, we pick up the residue at the leading pole $j=ix-\tau/4$ to get
\begin{align}\label{P-as}
P_{S,\tau}(x) = S^{-\tau/2+2ix} {\Gamma^2(\tau/2)\over \Gamma^2(\tau/4+ ix)} + \dots\,,
\end{align}
where the dots denote subleading terms.

Taking the sum over $S=0,1,\dots$ on both sides of \re{recrel}  we get
\begin{align}
2 i x\sum_{S\ge 0} (2 S+\tau -1)P_{S,\tau}(x) = (\tau/2 -2)\sum_{S\ge 0}  (2 S+\tau -1)P_{S,\tau}(x) -
 (\tau -2) (\tau/2 -1) 
\end{align}
wherefrom
\begin{align}\label{Sum1}
\sum_{S\ge 0} (2 S+\tau -1)P_{S,\tau}(x) = {(\tau -2)^2 \over 2(\tau/2 -2-2 i x)}\,.
\end{align}
Then, we multiply both sides of  \re{recrel} by $\psi(S+\tau/2)-\psi(1)$, with $\psi(x)= (\ln \Gamma(x))'$ being the Euler $\psi-$function, and go through the same steps to get
\begin{align}\notag
(\tau/2 -2-2 i x) {}& \sum_{S\ge 0} (2 S+\tau -1)\left[\psi(S+\tau/2)-\psi(1)\right]P_{S,\tau}(x)  
\\
{}& =\sum_{S\ge 0} (2 S+\tau -1)P_{S,\tau}(x) + \frac12(\tau-2)^2\left[\psi(\tau/2-1)-\psi(1)\right]\,.
\end{align}
Together with \re{Sum1} this leads to
\begin{align}\notag
 \sum_{S\ge 0}{}& (2 S+\tau -1)\left[\psi(S+\tau/2)-\psi(1)\right]P_{S,\tau}(x) 
 \\
 {}& = {(\tau -2)^2 \over 2(\tau/2 -2-2 i x)^2} +  {(\tau-2)^2\over 2(\tau/2 -2-2 i x)}\left[\psi(\tau/2-1)-\psi(1)\right]\,.
\end{align}
Replacing $\tau=6$ we encounter the sum that enters \re{mech1}.

\bibliographystyle{JHEP}

\providecommand{\href}[2]{#2}\begingroup\raggedright\endgroup
 
\end{document}